\documentstyle[12pt]{article}
\newcommand\bea{\begin{eqnarray}}
\newcommand\eea{\end{eqnarray}}
\begin{document}
\bibliographystyle{unsrt}
\setlength{\baselineskip}{18pt}
\parindent 18pt

\begin{center}{
{\Large {\bf Deformation of quantum oscillator and of its interaction with
environment}} }
\vskip 1truecm
A. Isar${\dagger\ddagger}^{(a)}$ and W. Scheid ${\ddagger}$\\
$\dagger${\it Department of Theoretical Physics, Institute of Physics and
Nuclear Engineering\\
Bucharest-Magurele, Romania }\\
$\ddagger${\it Institut f\"ur Theoretische Physik der
Justus-Liebig-Universit\"at \\ Giessen, Germany }\\
\end{center}

\begin{abstract}
A master equation for the deformed quantum harmonic oscillator
interacting with a dissipative environment, in particular with a thermal
bath, is derived in the microscopic model by using
perturbation theory, for the case when the interaction is deformed. The
coefficients of the master equation and of equations
of motion for observables depend on the deformation function. The steady
state solution of the equation for the density matrix in the number
representation is obtained and shown that it satisfies the detailed balance
condition. The equilibrium energy of the deformed harmonic oscillator,
calculated in the
approximation of small deformation, does not depend on the deformation of
interaction operators.
\end{abstract}


\vskip 0.5truecm (a) e-mail address: isar@theory.nipne.ro

\section{Introduction}

For more than a decade a constant interest has been manifested to
the study of deformations of Lie algebras, so-called  quantum algebras or
quantum groups, whose rich structure produced important
results and consequences in statistical mechanics,
quantum field theory, conformal field theory, quantum and nonlinear optics,
nuclear and
molecular physics \cite{ray,bon}.
Their use in physics became stronger with the introduction
of the  $q$-deformed Heisenberg-Weyl algebra ($q$-deformed
quantum harmonic oscillator) by Biedenharn \cite{bied} and MacFarlane
\cite{macf} in 1989.
There are, at least, two properties which make $q$-oscillators interesting
objects for physics. The first is the fact that they naturally appear as the
basic building blocks
of completely integrable theories. The second concerns the
connection between $q$-deformation and nonlinearity.
In Refs. \cite{mank2,mank1,ani} it was shown that the $q$-oscillator leads to
nonlinear vibrations with a special kind of the dependence of the
frequency on the amplitude. The $q$-deformed Bose distribution has been
obtained
and it  was also shown how $q$-nonlinearity produces a correction to the Planck
distribution formula \cite{mank2,mank1,su}.

The present paper is the third one in the series of papers devoted to the
study of the influence of quantum
deformation on quantum dissipation. In Ref. \cite{adef1}, using a
variant of Mancini's model \cite{manc}, we derived a master equation for
the $f$-deformed
oscillator in the presence of a dissipative environment for an undeformed
interaction between system and environment. Then in Ref.
\cite{adef2} we obtained a Lindblad master equation for the ordinary
harmonic oscillator interacting with an environment through a
deformed interaction. The equations of motion obtained for different
observables had a strong dependence on the deformation.
In the present paper we set a master equation
for the deformed harmonic oscillator in the presence of a dissipative
environment, for the case of a deformed interaction of the system with
its environment. This equation is shown to be
a deformed version of the master equation obtained in the framework
of the Lindblad theory for open quantum systems \cite{l1}.
When the deformation becomes zero, we recover the
Lindblad master equation for the damped harmonic oscillator \cite{ss,rev}.
We are interested in studying
the role of nonlinearities which appear in the master equation.
This goal is motivated by the fact
that the $q$-oscillator can be considered as a physical system with a
specific nonlinearity, called $q$-nonlinearity \cite{mank2,mank1}.
For a certain choice of the environment coefficients, a
master equation for the damped deformed oscillator has also been derived by
Mancini \cite{manc}.

The paper is organized as follows. In Sec. 2 we remind the
basics of the $f$-deformed quantum oscillator, in particular of the
$q$-oscillator. In Sec. 3 we derive a master equation for the $f$-deformed
oscillator
in the presence of a dissipative environment, for the case of a deformed
interaction between system and environment.
The equations of motion obtained for different observables
present a strong dependence on the deformation.
Then in Sec. 4 we give
the equation for the density matrix in the number representation and
find the stationary state.
In the particular case when the environment is a thermal
bath we obtain an expression for the equilibrium energy of the oscillator
in the approximation of a small deformation parameter.
A summary and conclusions are given in Sec. 5.

\section{$f$- and $q$-deformed quantum oscillators}

It is known that the ordinary operators $\{1, a, a^\dagger, N\}$ form the Lie
algebra of the Heisenberg-Weyl group and the linear harmonic oscillator can be
connected with the generators of the Heisenberg-Weyl Lie group.
The $f$-deformed quantum oscillators \cite{mank3}
are defined by the algebra
generated by the operators $\{1,A,A^\dagger,N\},$ where
the Hermitian number operator $N$ is not equal to $A^\dagger A$ as in the
ordinary case.

The $f$-deformed oscillator operators are given as follows
\cite{mank3}: \bea
A=af(N)=f(N+1)a,~~A^\dagger=f(N)a^\dagger=a^\dagger
f(N+1),\label{def}\eea where $N=a^\dagger a.$ They satisfy the
commutation relations \bea [A,N]=A,~~[A^\dagger,N]=-A^\dagger
\label{com1}\eea and \bea
[A,A^\dagger]=(N+1)f^2(N+1)-Nf^2(N).\label{com3}\eea The function
$f,$ which is a characteristics for the deformation, has a
dependence on a deformation parameter $\alpha$ such that when the
deformation disappears, then $f(N,\alpha=0)=1$ and the usual algebra
is recovered. Transformation (\ref{def}) of the operators
$a,a^\dagger$ to $A,A^\dagger$ represents a nonlinear noncanonical
transformation, since it does not preserve the commutation relation,
i. e. $[A,A^\dagger]\not= 1.$

The notion of $f$-oscillators generalizes the notion of $q$-oscillators
($q=1+\alpha$). Indeed, for
\bea f(N)=\sqrt{[N]\over N},~~[N]\equiv{q^N-q^{-N}\over q-q^{-1}},
\label{qdef}\eea
where $q$ is the deformation parameter (a dimensionless $c$-number),
the operators $A$ and $A^\dagger$ in Eq. (\ref{com1}) become the $q$-deformed boson
annihilation and creation operators \cite{bied,macf}. For
the $q$-deformed harmonic oscillator the commutation relation (\ref{com3}) becomes
\bea [A,A^\dagger]=[N+1]-[N].\label{com2}\eea
If $q$ is real positive, then
\bea [N]={\sinh(N\ln q)\over \sinh(\ln q)}\eea
and the condition of Hermitian conjugation
$(A^\dagger)^\dagger=A$ is satisfied.
In the limit $q\to 1,$ $q$-deformed operators
tend to the ordinary operators because
$\lim_{q\to 1} [N]=N.$
Then Eq. (\ref{com2}) goes to the usual boson
commutation relation $[A,A^\dagger]=1.$

The $q$-deformed boson operators $A$ and $A^\dagger$ can be expressed in terms
of the usual boson operators $a$ and $a^\dagger$ (satisfying $[a,a^\dagger]=1,
$ $N=a^\dagger a$ and
$[a,N]=a,~~[a^\dagger,N]=-a^\dagger$) through the relations
\cite{kuli,song} (see Eqs. (\ref{def})):
\bea A=a\sqrt{[N]\over N}=\sqrt{[N+1]\over N+1}a,~~
A^\dagger=\sqrt{[N]\over N}a^\dagger=a^\dagger  \sqrt{[N+1]\over N+1}.\eea

Using the nonlinear map (\ref{qdef}) \cite{poly,curt}, the $q$-oscillator has been
interpreted
\cite{mank2,mank1} as a nonlinear oscillator with a special type of
nonlinearity which
classically corresponds to an energy dependence of the oscillator frequency.
Other nonlinearities can also be introduced by making the frequency
to depend on other constants of motion,
different from energy \cite{mank2,mank3}, through the deformation function $f.$
Other examples of deformation functions $f$ can be found in Refs.
\cite{mat,sud,aga,dod}.

The Hamiltonian of the $f$-deformed harmonic oscillator ($\omega$ is
the ordinary frequency) is a function of $N$: \bea {\cal
H}={\hbar\omega\over 2}(AA^\dagger+A^\dagger A) ={\hbar\omega\over
2}\left(\left(N+1\right)f^2\left(N+1\right)+Nf^2\left(N\right)\right).\label{ham}\eea
It is diagonal on the eigenstates  $|n>$ in the Fock space and its
eigenvalues are \bea E_n={\hbar\omega\over
2}\left(\left(n+1\right)f^2\left(n+1\right)+nf^2\left(n\right)\right).\label
{eig}\eea In the limit $f\to 1$ ($q\to 1$ for $q$-oscillators), we
recover the ordinary expression $E_n=\hbar\omega(n+1/2).$

Using the operator Heisenberg equation with Hamiltonian $\cal $
(\ref{ham}) \bea i\hbar{dA\over dt}=[{A,\cal H}]\eea we obtain the
following solutions to the Heisenberg equations of motion for the
$f$-deformed operators $A$ and $A^\dagger $ defined in Eqs.
(\ref{def}) \cite{manc,mank3}: \bea
A(t)=\exp\left(-i\omega\Omega\left(N\right)t\right)A,~~A^\dagger(t)=A^\dagger
\exp\left(i\omega \Omega\left(N\right)t\right),\label{temp}\eea
where $\Omega(N)$ is the operator defined as \bea \Omega(N)={1\over
2}\left(\left(N+2\right)f^2\left(N+2\right)-Nf^2\left(N\right)\right).\eea
For a $q$-deformed harmonic oscillator, \bea \Omega(N)={1\over
2}([N+2]-[N])\label{qom}\eea and for a small deformation parameter
$\tau$ ($\tau=\ln q$), \bea \Omega(N)=1+{\tau^2\over
2}(N+1)^2.\label{deform}\eea

\section{Quantum Markovian master equation}

In order to discuss the dynamics of the open systems S, we use a
microscopic description of the composite system. As the subsystem S
of interest we take the $f$-deformed  harmonic oscillator with
Hamiltonian $\cal H$ (\ref{ham}) and the environment R (reservoir,
bath) with the Hamiltonian $H_R.$ The coupled system S+R with the
total Hamiltonian $H_T={\cal H}+H_R+{\cal V}$ ($\cal V$ is the
$f$-deformed interaction Hamiltonian) is described by a density
operator $\chi(t),$ which evolves in time according to the von
Neumann-Liouville equation \bea {d\chi(t)\over dt}=-{i\over
\hbar}[H_T,\chi(t)].\eea When the Hamiltonian evolution of the total
system is projected onto the space of the harmonic oscillator, the
reduced density operator of the subsystem is given by $\rho(t)={\rm
Tr_R}\chi(t).$ The derivation of the reduced density operator in
which the operators of the environment system have been eliminated
up to second order of the perturbation theory can be taken from
literature \cite{loui,hake,gard,carm,wal}. Following \cite{carm} and
the procedure used in the previous paper \cite{adef1}, we point the
main steps in obtaining the master equation which describes the time
evolution of the damped deformed harmonic oscillator. We assume that
the initial state $R_0$ of the environment at $t=0$ does not depend
on the state $\rho(0)$ of the subsystem. Then $\chi(0)=\rho(0)R_0.$
At later times correlations between S and R arise due to the
coupling of the system and environment through $\cal V.$ However, we
assume that this coupling is weak and then at any instant of time,
not only for the initial time, the reservoir and the oscillator are
approximately decoupled. Furthermore, R is a large system whose
state should be virtually unaffected by its coupling to S. We then
write $\chi(t)=\rho(t)R_0.$ In the previous paper \cite{adef1}, we
considered the interaction potential of the linear form in the
coordinate and momentum operators in the Hilbert space of the
subsystem S. In analogy to that model, we assume in the present
paper a $f$-deformed interaction potential $\cal V$ of the form \bea
{\cal V}=\sum_{i=1,2} S_i \Gamma_i, \label{int}\eea where $S_1=Q$
and $S_2=P$ are the $f$-deformed coordinate and momentum operators
in the Hilbert space of the subsystem S ($m$ is the oscillator
mass): \bea Q=\sqrt{\hbar\over 2m\omega}(A^\dagger+A), ~~
P=i\sqrt{\hbar m\omega\over 2}(A^\dagger-A), \eea with $A^\dagger,A$
defined in Eqs. (\ref{def}) and $\Gamma_i$ are Hermitian operators
in the Hilbert space of the environment. Then we obtain the
following master equation for the density operator of the open
quantum system in the Born-Markov approximation:
\bea {d\rho(t)\over dt}=-{i\over \hbar}[{\cal H},\rho(t)]
+{1\over\hbar^2}\sum_{i,j=1,2}\int_0^t dt'\{
C_{ij}^*(t')[S_i,\rho(t)S_j(-t')]
+C_{ij}(t')[S_j(-t')\rho(t),S_i]\},\label{mast}\eea where the
coefficients $C_{ij}(t')={\rm Tr}_R\{R_0\Gamma_i(t')\Gamma_j\}$ are
correlation functions of the environment operators. It is assumed
that the correlation functions decay very rapidly on the time scale
on which $\rho(t)$ varies. Ideally, we might take $C_{ij}(t')\sim
\delta(t').$ The Markov approximation relies on the existence of two
widely separated time scales: a slow time scale for the dynamics of
the system S and a fast time scale characterizing the decay of
environment correlation functions \cite{carm}.

In order to get the time dependence of the operators of
coordinate $S_1(t)=Q(t)$ and momentum $S_2(t)=P(t)$ used in Eq. (\ref{mast}),
we express them through the relations
\bea Q(t)=\sqrt{\hbar\over 2m\omega}(A^\dagger(t)+A(t)), ~~
P(t)=i\sqrt{\hbar m\omega\over 2}(A^\dagger(t)-A(t)) \eea
and then insert Eqs. (\ref{temp}) for $A(t)$ and $A^\dagger(t).$ Then the
master equation (\ref{mast}) takes the following form:
\bea {d\rho(t)\over dt}=
-{i\over \hbar}[{\cal H},\rho(t)]\nonumber\\
+{1\over 2\hbar^2}\int_0^t dt'
(C_{11}^*(t')[Q,\rho(t)\left(QE_-+E_+Q-{i\over m\omega}\left(PE_--E_+P\right)\right)]
\nonumber\\
+C_{11}(t')[\left(QE_-+E_+Q-{i\over m\omega}\left(PE_--E_+P\right)\right)\rho(t),Q]
\nonumber\\
+iC_{22}^*(t')[P,\rho(t)\bigl( m\omega(QE_--E_+Q)-i(PE_-+E_+P)\bigr)]\nonumber\\
+iC_{22}(t')[\bigl( m\omega\left(QE_--E_+Q\right)
-i\left(PE_-+E_+P\right)\bigr)\rho(t),P]\nonumber\\
+iC_{12}^*(t')[Q,\rho(t)\bigl( m\omega\left(QE_--E_+Q\right)
-i\left(PE_-+E_+P\right)\bigr)]\nonumber\\
+C_{21}(t')[\left(QE_-+E_+Q-{i\over m\omega}\left(PE_--E_+P\right)\right)\rho(t),P]
\nonumber\\
+C_{21}^*(t')[P,\rho(t)\left(QE_-+E_+Q-{i\over m\omega}\left(PE_--E_+P\right)\right)]
\nonumber\\
+iC_{12}(t')[\bigl( m\omega\left(QE_--E_+Q\right)
-i\left(PE_-+E_+P\right)\bigr)\rho(t),Q]),
\label{mast1}\eea
where  we have introduced the following notations:
\bea E_+=\exp\bigl(i\omega\Omega\left(N\right)t'\bigr),
~~E_-=\exp\left(-i\omega\Omega\left(N\right)t'\right).\eea

If the environment is sufficiently large we may assume that the time
correlation functions decay fast enough to zero for times $t'$
longer than the relaxation time $t_B$ of the environment: $t'\gg
t_B.$ Therefore, if we are interested in the dynamics of the
subsystem over times which are longer than the environment
relaxation time,  $t\gg t_B,$ we may use the Markov approximation
and replace the upper limit of integration $t$ by $\infty.$
Physically, this amounts to assuming that the memory functions
$C_{ij}(t')E_{\pm}(t')$ decay over a time which is much shorter than
the characteristic evolution time of the system of interest. We
evaluate the integrals in Eq. (\ref{mast1}) like in the preceding
paper \cite{adef1}. After certain assumptions \cite{loui,wal}, one
can define the complex decay rates, which govern the rate of
relaxation of the system density operator, as follows: \bea
\int_0^\infty dt' C_{11}(t')E_{\pm}= \int_0^\infty dt'
C_{11}^*(t')E_{\pm}=D_{pp}(\Omega),\label{dif1}\eea \bea
\int_0^\infty dt'C_{22}(t')E_{\pm}= \int_0^\infty
dt'C_{22}^*(t')E_{\pm}=D_{qq}(\Omega),\eea \bea \int_0^\infty
dt'C_{12}(t')E_{\pm}=\int_0^\infty dt'C_{21}^*(t')E_{\pm}
=-D_{pq}(\Omega)+{i\hbar\over 2}\lambda(\Omega).\label{disco}\eea In
fact,  $D_{pp}(\Omega),$ $D_{qq}(\Omega),$ $D_{pq}(\Omega)$ and
$\lambda(\Omega)$ play the role of deformed diffusion and,
respectively,  dissipation coefficients. The existence of these
coefficients reflects the fact that, due to the interaction, the
energy of the system is dissipated into the environment, but noise
arises also (in particular, thermal noise), since the environment
also distributes some of its energy back to the system. Then the
master equation (\ref{mast1}) for the damped deformed harmonic
oscillator takes the form:
\bea   {d    \rho \over dt}=-{i\over \hbar}[{\cal H},    \rho]\nonumber\\
+{1\over 2\hbar^2}([\left(\{D_{pp}\left(\Omega\right),Q\}
+{i\over m\omega}[D_{pp}\left(\Omega\right),P]\right)
\rho,Q]+[\left(\{D_{qq}(\Omega),P\}-im\omega[D_{qq}(\Omega),Q]
\right)\rho,P]\nonumber\\
+[\left(m\omega[iD_{pq}\left(\Omega\right)+{\hbar\over 2}\lambda(\Omega),Q]
-\{D_{pq}\left(\Omega\right)-{i\hbar\over 2}
\lambda(\Omega),P\}\right)\rho,Q]\nonumber\\
-[\left({1\over m\omega}[iD_{pq}\left(\Omega\right)-{\hbar\over 2}
\lambda(\Omega),P]
+\{D_{pq}\left(\Omega\right)+{i\hbar\over 2}\lambda(\Omega),Q\}\right)\rho,P]
+H.c.).\label{mast2}\eea
Here the curly parentheses mean anticommutators. We notice that the deformation
is present in both the commutator
containing the oscillator Hamiltonian $\cal H,$ as well as in
the dissipative part of the master equation, which describes the influence of
the environment on the deformed oscillator.
This master equation preserves the Hermiticity property of the density
operator and the normalization (unit trace)
at all times, if at the initial time it has these properties.
In the limit $f\to 1$ $(\Omega\to 1),$ the deformation disappears and Eq.
(\ref{mast2})
becomes the Markovian master equation for the damped harmonic oscillator,
obtained in the Lindblad theory for open quantum systems, based on completely
positive dynamical semigroups \cite{l1,ss,rev}.
The fact that we introduced a $f$-deformed interaction between the open sytem
and environment is reflected in the presence of deformed operators $P$ and
$Q$ in the dissipative part of Eqs. (\ref{mast1}) and (\ref{mast2}), along with the
deformed diffusion operator coefficients, compared to the previous  model
\cite{adef1,manc}, where the dissipative part of the master equation contains
the deformed diffusion coefficients and the usual operators $p$ and $q.$

Expressing the coordinate and momentum operators back in terms of the
creation and annihilation operators, introducing the notations
\bea D_+(\Omega)\equiv {1\over 2\hbar}\left(m\omega
D_{qq}\left(\Omega\right)+{D_{pp}\left(\Omega\right)
\over m\omega}\right),~~D_-(\Omega)\equiv {1\over 2\hbar}\left(m\omega
D_{qq}\left(\Omega\right)-{D_{pp}\left(\Omega\right)\over m\omega}\right)\eea
and assuming, like in Ref. \cite{adef1}, that $\lambda(\Omega)=\lambda=const,$
the master equation (\ref{mast2}) for the damped deformed harmonic oscillator takes
the form
\bea   {d    \rho \over dt}=-{i\over \hbar}[{\cal H},    \rho]
+\bigl([[D_+(\Omega)af(N),\rho],f(N)a^\dagger]\nonumber\\-[[f(N)a^\dagger
\left(D_-\left(\Omega\right)
+{i\over\hbar}D_{pq}\left(\Omega\right)\right),\rho],f(N)a^\dagger]
-{\lambda\over 2}[f(N)a^\dagger,\{af(N),\rho\}]+H.c.\bigr),\label{mast5}\eea
with $\cal H$ given by Eq. (\ref{ham}).

In the particular case of a thermal equilibrium of the bath at temperature $T$
($k$ is the Boltzmann constant), we take the diffusion coefficients of the form
(in concordance with results of Refs. \cite{adef1,manc})
\bea m\omega D_{qq}(\Omega)={D_{pp}(\Omega)\over m\omega}=
{\hbar \over 2}\lambda\coth{\hbar\omega\Omega\over 2kT},
~~D_{pq}(\Omega)=0\label{defth}\eea
and the master equation (\ref{mast5}) takes the form \bea
 {d\rho \over dt}=-{i\over \hbar}[{\cal H},    \rho]
+{\lambda\over 2}\left([[\coth{\hbar\omega\Omega\over 2kT}af(N),\rho],
f(N)a^\dagger]-[f(N)a^\dagger,\{af(N),\rho\}]+H.c.\right).
\label{mast7}\eea
In the limit $\Omega\to 1,$ the deformed diffusion coefficients
(\ref{defth}) take the known form obtained for
the damped harmonic oscillator,
if the asymptotic state is a Gibbs state \cite{ss,rev}:
\bea D_{pp}={\hbar m\omega\over 2}\lambda\coth{\hbar\omega\over 2kT},
~~D_{qq}={\hbar\over 2m\omega}\lambda\coth{\hbar\omega\over 2kT},
~~D_{pq}=0.\label{coegib} \eea
If the bath temperature is $T=0,$ the master equation (\ref{mast7})
simplifies:
\bea   {d    \rho \over dt}=-{i\omega\over 2}
[(N+1)f^2(N+1)+Nf^2(N),\rho] \nonumber\\
-\lambda\left(Nf^2(N)\rho+\rho Nf^2(N)-2f(N+1)a\rho a^\dagger
f(N+1)\right).\label{mast6}\eea

The meaning of the master equation (\ref{mast7}) becomes clear when we transform
it into equations satisfied by the expectation values of
observables involved in the master equation, $<O>={\rm Tr}[\rho(t)O],$ where
$O$ is the operator corresponding to such an observable.
We give an example, multiplying both sides of Eq. (\ref{mast7})
by the number operator $N$ and taking the trace.
Then we obtain the following equation of motion for the expectation
value of $N$:\bea{d\over dt}<N>
=\lambda(<(\coth{\hbar\omega\Omega(N)\over 2kT}-1)(N+1)f^2(N+1)>\nonumber\\
-<(\coth{\hbar\omega\Omega(N-1)\over 2kT}+1)Nf^2(N)>).\label{expv}\eea
This equation leads to a time dependence of the averaged number of quanta on
dissipation, temperature and deformation, compared to the case of an oscillator
without dissipation, where the expectation number of quanta is conserved.
In the case of a $q$-deformation, Eq. (\ref{expv}) takes the form
\bea{d\over dt}<N>
=\lambda\left(<(\coth{\hbar\omega\Omega(N)\over 2kT}-1)[N+1]>
-<(\coth{\hbar\omega\Omega(N-1)\over 2kT}+1)[N]>\right).\label{expv1}\eea
If, in addition, the temperature of the thermal bath is $T=0,$
Eq. (\ref{expv1}) becomes
\bea{d\over dt}<N>=-2\lambda<[N]>.\label{expnum}\eea
In order to obtain an approximate solution of this equation, we work in the limit
of a small deformation parameter $\tau=\ln q$ ($q$ real).
Then we can take \cite{ray,bon}
\bea [N]=N+{\tau^2\over 6}(N^3-N)\eea
and, making also the assumption $<N^3>\approx<N>^3,$ Eq. (\ref{expnum}) reduces
to the following differential equation:
\bea {d\over dt}<N>=-2\lambda\left(<N>+{\tau^2\over 6}(<N>^3-<N>)\right).\eea
By integrating this equation we obtain
\bea {<N(t)>\over\sqrt{1-{\tau^2\over 6}+{\tau^2\over 6}
<N(t)>^2}}={<N(0)>e^{-2\lambda(1-{\tau^2\over 6})t}
\over\sqrt{1-{\tau^2\over 6}+{\tau^2\over 6} <N(0)>^2}},\eea
from where we can obtain the following expression for the expectation value of
the number operator in the approximation of a small deformation parameter:
\bea <N(t)>=\frac
{<N(0)>\sqrt{1-{\tau^2\over 6}}e^{-2\lambda(1-{\tau^2\over6})t}}
{\sqrt{1-{\tau^2\over 6}+{\tau^2\over 6}<N(0)>^2
(1-e^{-4\lambda(1-{\tau^2\over 6})t})}}\nonumber\\
\approx <N(0)>e^{-2\lambda(1-{\tau^2\over 6})t}
\left(1-{\tau^2\over 12}<N(0)>^2(1-e^{-4\lambda(1-{\tau^2\over 6})t})\right).\eea
In the limit $\tau\to 0,$ we obtain
\bea <N(t)>=<N(0)>e^{-2\lambda t},\eea
which is the expression of the expectation value of the number operator of the
damped harmonic oscillator obtained in the Lindblad theory for open quantum systems.

We consider another example, taking the simplest case of a thermal bath
at $T=0,$ when both diffusion and dissipation coefficients do
not depend on the deformation, $D_+=\lambda/2=const.$
Even in this situation, the equations of motion for the
expectation values are yet complicated, because they do not form a closed system.
Multiplying both sides of Eq. (\ref{mast6}) by the operator $a$
and, respectively, $\Omega(N)a$
and taking throughout the
trace, we get the following equations for the expectation values
of these operators:
\bea{d\over dt}<a>=-i\omega<\Omega(N)a>\nonumber\\
-\lambda<\left((N+1)f^2(N+1)+Nf^2(N)-2Nf(N)f(N+1)\right)a >,\eea
\bea{d\over dt}<\Omega(N)a>=-i\omega<\Omega^2(N)a> \nonumber \\
-\lambda<\left(\Omega(N)\left(\left(N+1\right)f^2\left(N+1\right)+Nf^2\left(N
\right)\right)-2\Omega(N-1)Nf(N)f(N+1)\right)a>.\eea
These examples show that the equations of motion contain strong
nonlinearities introduced by the deformation function $f$ and they do not form a
closed system of equations.

\section{Equation for the density matrix}

Let us rewrite the master equation (\ref{mast5}) for the density matrix by
means of the number representation. Namely, we take the matrix elements
of each term between different number states denoted by $|n>$, and using
$N|n>=n|n>,$ $a^+|n>=\sqrt{n+1}|n+1>$ and $a| n>=\sqrt n| n-1>,$ we get
\bea {d\rho_{mn}\over dt}=-{i\omega\over 2}[mf^2(m)+(m+1)f^2(m+1)
-nf^2(n)-(n+1)f^2(n+1)]\rho_{mn}\nonumber\\
-[(m+1)f^2(m+1)\left(D_+(\Omega(m))-{\lambda\over 2}
\right)+mf^2(m)\left(D_+(\Omega(m-1))+{\lambda\over 2}
\right)\nonumber \\+(n+1)f^2(n+1)\left(D_+(\Omega(n))-{\lambda\over 2}\right)+nf^2(n)
\left(D_+(\Omega(n-1))+{\lambda\over 2}\right)]\rho_{mn}\nonumber\\
+\sqrt{(m+1)(n+1)}f(m+1)f(n+1)[D_+(\Omega(m))+D_+(\Omega(n))+\lambda]
\rho_{m+1,n+1}\nonumber\\
+\sqrt{mn}f(m)f(n)[D_+(\Omega(m-1))+D_+(\Omega(n-1))-\lambda]\rho_{m-1,n-1}
\nonumber\\
-\sqrt{(m+1)n}f(m+1)f^2(n)[D_-(\Omega(m))+D_-(\Omega(n-1))\nonumber\\-{i\over
\hbar}\left(D_{pq}(\Omega(m))
+D_{pq}(\Omega(n-1))\right)]\rho_{m+1,n-1}\nonumber\\
-\sqrt{m(n+1)}f(m)f(n+1)[D_-(\Omega(m-1))+D_-(\Omega(n))\nonumber\\
+{i\over\hbar}\left(D_{pq}
(\Omega(m-1))+D_{pq}(\Omega(n))\right)]\rho_{m-1,n+1}\nonumber\\
+\sqrt{(m+1)(m+2)}f(m+1)f(m+2)[D_-(\Omega(m+1))-{i\over \hbar}D_{pq}(\Omega(m+1))]
\rho_{m+2,n}
\nonumber\\
+\sqrt{(n+1)(n+2)}f(n+1)f(n+2)[D_-(\Omega(n+1))+{i\over \hbar}D_{pq}(\Omega(n+1))]
\rho_{m,n+2}\nonumber\\
+\sqrt{m(m-1)}f(m-1)f(m)[D_-(\Omega(m-2))+{i\over \hbar}D_{pq}(\Omega(m-2))]
\rho_{m-2,n}
\nonumber\\
+\sqrt{n(n-1)}f(n-1)f(n)[D_-(\Omega(n-2))-{i\over \hbar}D_{pq}(\Omega(n-2))]
\rho_{m,n-2}.
\label{numeq}\eea
Here we used the abbreviated notation $\rho_{mn}=<m\vert\rho(t)\vert n>.$
This equation, very complicated in form and in indices involved,
gives an infinite hierarchy of coupled
equations for the matrix elements. When
\bea D_-(\Omega(n))=0,~~D_{pq}(\Omega(n))=0,\eea
the diagonal elements are coupled only amongst themselves and not coupled to
the off-diagonal elements. In this case the diagonal elements
(populations) satisfy a simpler set of master equations:
\bea {dP(n)\over dt}=-\left((n+1)f^2(n+1)\left(2D_+(\Omega(n))
-\lambda\right)+nf^2(n)\left(2D_+(
\Omega(n-1))+\lambda\right)\right)P(n)\nonumber\\
+(n+1)f^2(n+1)\left(2D_+(\Omega(n))+\lambda\right)P(n+1)
+nf^2(n)\left(2D_+(\Omega(n-1))-\lambda\right)P(n-1),\label{pop}\eea
where we have set $P(n)\equiv\rho_{nn}.$
For a $q$-deformation, Eq. (\ref{pop}) takes the form \bea {dP(n)
\over dt}=-\left([n+1]\left(2D_+(\Omega(n))
-\lambda\right)+[n]\left(2D_+(\Omega(n-1))+
\lambda\right)\right)P(n)\nonumber\\
+[n+1]\left(2D_+(\Omega(n))+\lambda\right)P(n+1)
+[n]\left(2D_+(\Omega(n-1))-\lambda\right)P(n-1),\label{pop2}\eea
where, according to Eq. (\ref{qom}), \bea \Omega(n)={1\over
2}([n+2]-[n]).\eea We define the transition rates \bea
t_+(n)=(n+1)f^2(n+1)\left(2D_+(\Omega(n))-\lambda\right),
~~t_-(n)=nf^2(n)\left(2D_+(\Omega(n-1))+\lambda\right),\eea which
for a $q$-deformation look like \bea
t_+(n)=[n+1]\left(2D_+(\Omega(n))-\lambda\right),
~~t_-(n)=[n]\left(2D_+(\Omega(n-1))+\lambda\right).\eea With these
notations Eq. (\ref{pop}) becomes: \bea {dP(n)\over
dt}=t_+(n-1)P(n-1)+t_-(n+1)P(n+1)-\left(t_+(n)+t_-(n)\right)P(n)
\label{pop1}.\eea The steady state solution of Eq. (\ref{pop1}) is
found to be \bea P_{ss}(n)=P(0)\prod_{k=1}^{n}{2D_+(\Omega(k-1))-
\lambda\over 2D_+(\Omega(k-1))+\lambda}.\label{stst}\eea We remark
that in the steady state the detailed balance condition holds: \bea
t_-(n)P(n)=t_+(n-1)P(n-1).\eea In the particular case of a thermal
state, when the diffusion coefficients have the form (\ref{defth}),
the stationary solution of Eq. (\ref{pop1}) takes the following
form: \bea P_{ss}^{th}(n)=Z_f^{-1}\exp\{-{\hbar\omega\over 2 kT}
\left((n+1)f^2(n+1)+nf^2(n)\right)\}\label{popth},\eea where \bea
Z_f^{-1}=P(0)\exp{\hbar\omega f^2(1)\over 2kT}\eea and $Z_f$ is the
partition function: \bea
Z_f=\sum_{n=0}^{\infty}\exp\{-{\hbar\omega\over 2 kT}
\left((n+1)f^2(n+1)+nf^2(n)\right)\}\label{part}.\eea By using the
eigenvalues (\ref{eig}), the distribution (\ref{popth}) can be
written \bea P_{ss}^{th}(n)=Z_f^{-1}\exp(-{E_n\over
kT}).\label{sten}\eea Expressions (\ref{popth}) and (\ref{sten})
represent the Boltzmann distribution for the deformed harmonic
oscillator. In the limit $f\to 1$ the probability $P_{ss}^{th}(n)$
becomes the usual Boltzmann distribution for the ordinary harmonic
oscillator with the well-known partition function
$Z=1/2\sinh{\hbar\omega\over 2kT}.$ For the $q$-oscillator described
by Hamiltonian $\cal H$ in Eq. (\ref{ham}) and weakly coupled to a
reservoir kept at the temperature $T,$ the $q$-deformed partition
function can be obtained as a particular case of the partition
function $Z_f$ (\ref{part}), by taking the deformation function
(\ref{qdef}): \bea Z_q=\sum_{n=0}^\infty\exp\{-{\hbar\omega\over
2kT}{\sinh(\tau(n+1))+ \sinh(\tau n)\over \sinh
\tau}\}.\label{qpart}\eea The results (\ref{stst}) -- (\ref{qpart})
coincide with those obtained in the previous paper \cite {adef1},
where we considered a linear interaction potential $V$ and the
harmonic oscillator operators contained in this potential are kept
undeformed. Therefore, for the equilibrium energy we obtain in the
limit of small deformation parameter $\tau$ the same expression like
that obtained in Ref. \cite{adef1}: \bea
E(t\to\infty)={\hbar\omega\over 2}(\coth{\hbar\omega\over
2kT}+{\tau^2}c),\eea where \bea c={e^{\beta}\over
(e^\beta-1)^2}\left({e^\beta+1\over e^\beta-1}
-\beta{e^{2\beta}+4e^\beta+1\over (e^\beta-1)^2}\right),
~~\beta={\hbar\omega\over kT}.\eea We note that in the approximation
of small deformation, the energy of the damped deformed oscillator
depends on the energy $\hbar\omega/2$ of oscillator ground state and
on the temperature $T.$ Evidently, when there is no deformation
$(\tau\to 0),$ one recovers the energy of the ordinary harmonic
oscillator in a thermal bath \cite{ss,rev}. In the limit $T\to 0,$
one has $c\to 0,$ $E(\infty)=\hbar\omega/2$ and the deformation does
not play any role.

\section{Summary and conclusions}

Our purpose was to study the dynamics of the deformed quantum harmonic
oscillator in a deformed interaction with a dissipative environment, in particular
with a thermal bath. We derived in the Born-Markov approximation
a master equation for the reduced density
operator of the damped $f$-deformed oscillator.
The one-dimensional $f$- or $q$-oscillator is a nonlinear
quantum oscillator with a specific type of nonlinearity and,
consequently, the diffusion and dissipation coefficients
which model the
influence of the environment on the deformed oscillator strongly depend
on the introduced nonlinearities.
Compared to the previous paper \cite{adef1}, in the present work the harmonic
oscillator operators contained in the interaction terms are deformed. This fact is
reflected in the increased degree of the nonlinearity of the
equations of motion for the expectation values of observables.
In the limit of
zero deformation, the master equation takes the form of the
master equation for the damped oscillator obtained in the framework of the
Lindblad theory of open systems based on quantum dynamical semigroups.
We have also derived the equation for the density matrix in the number
representation. In the case of a thermal bath we obtained the stationary
solution, which is the Boltzmann distribution for the deformed harmonic oscillator.
In the steady state the detailed balance condition holds true.
In the approximation of a small deformation,
we obtained the expression of the equilibrium energy
of the deformed harmonic oscillator without deformation of interaction. This
energy depends on the oscillator ground state energy and temperature.

The master equation for the damped deformed harmonic oscillator is
an operator equation. It could be useful to study its consequences
for the density operator by transforming this equation into more
familiar forms, such as the partial differential equations of
Fokker-Planck type for the Glauber, antinormal ordering and Wigner
quasiprobability distributions or for analogous deformed
quasiprobabilities \cite{ani} associated with the density operator.
It could also be interesting to study the properties of the entropy
of the damped deformed harmonic oscillator and to find the states
which minimize the rate of entropy production for this system. In
the case of the undeformed damped oscillator such states are
represented by correlated coherent states \cite{apur}. We suppose
that for the damped deformed oscillator the corresponding states are
the deformed (nonlinear) coherent states \cite{mank3}, which play an
important role in the description of the phenomenon of environment
induced decoherence. The dissipative dynamics of deformed coherent
states superposition and the related coherence properties have
recently been studied by Mancini and Man'ko \cite{manc1}.

{\bf Acknowledgements}

One of us (A. I.) is pleased to express his sincere gratitude for
the hospitality at the Institut f\"ur Theoretische Physik in
Giessen. A. I. also gratefully acknowledges the fellowship from DAAD.

\end{document}